\documentstyle[aps,preprint,epsfig]{revtex}          
\tighten
\begin{document}
\title{The Yukawa Coupling in Three Dimensions}
\author{
F.A. Dilkes \\ 
D.G.C. McKeon \thanks{Tel. (519) 679-2111 x8789 \hfill Fax: (519) 661-3523 \hfill email: tmleafs@apmaths.uwo.ca}
\\ 
K. Nguyen }
\address{ Department 
of Applied Mathematics \\ University of Western Ontario \\ London, Ontario \\
CANADA \\ N6A 5B7}
\date{\verb$Date: 1997/05/26 19:27:51 $}

\maketitle

\begin{abstract}

We consider several renormalizable, scale free models in three
space-time dimensions which involve scalar
and spinor fields.  The Yukawa couplings are bilinear in both
the spinor and scalar fields and the potential is of 
sixth order in the scalar field.  In a model with a single scalar field and 
a complex Fermion field in three Euclidean dimensions, the couplings
in the theory are both asymptotically free.  This property is not retained
in $2+1$ dimensional Minkowski space, as we illustrate by considering
a renormalizable scale-free supersymmetric model.  This is on account of 
the different properties of the Dirac matrices in Euclidean and Minkowski
space.  We also examine a model in $2+1$ dimensional Minkowski space in which
two species of Fermions, associated with the two
unitarily inequivalent representations of the $2 \times 2$ Dirac
matrices, couple in two different ways to two distinct scalar fields.  
There are two types of Yukawa couplings in this model, 
and either one or the other of them can
be asymptotically free (but not both simultaneously).
 
\end{abstract}

\section{Introduction}

Self-interacting scalar models are the simplest of all relativistic quantum
theories to analyze.  There are but three instances in which the 
self-interaction involves a dimensionless coupling $\lambda$, (and 
is consequently perturbatively renormalizable): $\lambda \phi^4$ in
four dimensions, $\lambda \phi^3$ in six dimensions and
$\lambda \phi^6$ in three dimensions.  The 
renormalization group functions for these models have been determined
beyond one-loop order using dimensional regularization \cite{r1};
see refs.\cite{r2,r3,r4} respectively.  In this paper we consider
not just sixth order scalar couplings in three dimensions, but append
to them Yukawa couplings which are bilinear in both the scalar
and spinor fields.   This interaction is renormalizable in three
dimensions.

We compute the renormalization group functions associated with
these couplings to first order in the scalar self-coupling and second order
in the Yukawa coupling in a variety of models.  First, a simple model in which there is 
a single scalar field $\phi$ and a two-component Dirac Fermion field
$\psi$ is considered in Euclidean space and we find it to be asymptotically free.  
Next
a supersymmetric model in Minkowski space is examined
and, on account of the properties of Dirac matrices in this space, the
model no longer possesses asymptotic freedom.  Finally, a model containing
two species of two-component Dirac Fermions and two types of scalars, coupled
in a symmetric way that employs two distinct scalar self-couplings
and two Yukawa couplings, is analyzed.  It turns out  that it is
possible to arrange for one of the Yukawa couplings to be asymptotically free.

Dimensional regularization \cite{r1} is employed to compute
Feynman integrals for several reasons.  First of all, massless tadpoles
vanish when using this technique, eliminating a number of Feynman 
graphs.  Secondly, divergences in an odd number of dimensions occur
only beyond  one-loop order.  Thirdly, minimal subtraction \cite{r5}
is an efficient renormalization scheme which permits one to compute
the renormalization group functions with relative ease; this technique is 
contingent upon using dimensional regularization.  Fourthly, if one
uses dimensional regularization in a three dimensional model that initially
is free of dimensionful parameters, then one is not forced by consideration
of radiatively induced divergences to introduce couplings involving
massive parameters (such as masses for the scalar and spinor fields and
quartic scalar couplings).  We now consider the various models.

\section{The Yukawa Model}

Initially, let us consider the model defined by the three-dimensional
Euclidean action

\begin{equation} 
\label{e1}
S = \int d^3x \left[ -\case{1}{2}\phi p^2 \phi - \frac{\lambda}{6!}
\phi^6 - \overline \psi \left( p \!\! / + \case{1}{2} g \phi^2 \right)
\psi \right] \, . 
\end{equation}

The conventions used are given in the appendix.

In three dimensions, the couplings $\lambda$ and $g$ are dimensionless.
When dimensional regularization \cite{r1} is used, only these couplings
and the wave functions $\psi$ and $\phi$ need to be renormalized.  In 
computing the renormalization group functions, it is necessary to evaluate
the divergent parts of the two-point functions $\langle \phi \phi \rangle$,
and $\langle \psi \overline  \psi \rangle $, 
the four-point function 
$\langle \psi \overline
\psi \phi \phi \rangle$, 
and the six-point functions $\langle \phi \phi \phi \phi \phi \phi\rangle $.

As was mentioned in the introduction, using dimensional regularization with
this model means that tadpole and one-loop diagrams do not contribute to
the renormalization group functions.  Consequently, to order $\lambda$ and 
$g^2$, we only consider the diagrams in fig.\ref{f2}, using the Feynman 
rules of fig.\ref{f1}.  In table~\ref{t1} below, we list the number of 
distinct diagrams associated with each of the graphs in fig.\ref{f2}, the
associated combinatoric factors, and the 
pole part of the diagram (in $3- \epsilon$ dimensions).

The divergences computed in table~\ref{t1} do not contribute to 
vertices that occur in one-loop diagrams to the order which
we are considering.

Analogous calculations are easily performed in the supersymmetric version
of the model of eq.(\ref{e1}).  The Minkowski space superfield action for this
model is (using notation explained in the appendix)
\begin{equation}
\label{e2}
S = \int d^3x d^2 \theta \left( \case{1}{2} \Phi D^2 \Phi + \frac{\lambda}{4!}
\Phi^4 \right)
\end{equation}
which in component form becomes
\begin{eqnarray}
S &  = & \int d^3x \Big[ \case{1}{2} \left( A \Box A + i \psi^\alpha 
{\partial_\alpha}^\beta \psi_\beta + F^2 \right) \nonumber \\
& & \hspace{0.5cm} + \frac{\lambda}{4!} \left( 12 A^2 \psi^2 
+ 4 A^3 F \right) \Big] \, , \label{e3}
\end{eqnarray}
where $\psi$ is a Majorana two-component spinor in this theory; $A$ 
and $F$ are both real scalar fields.
The equation of motion for the auxiliary field $F$ may be used to re-express
(\ref{e3}) in the form
\begin{equation}
S = \int d^3 x \left[ \frac{1}{2} \left( A \Box A + i \psi^\alpha
{\partial_\alpha}^\beta \psi_\beta \right) + \frac{\lambda}{2}
A^2 \psi^2 - \frac{\lambda^2}{72} A^6 \right]
\label{e4}
\end{equation}
which is closely related to the action of (\ref{e1}).

We determine the lowest order radiative corrections to the coupling $\lambda$
by computing the divergent parts of the integrals corresponding to the 
diagrams in figures 
\ref{f2}c,
\ref{f2}d,
\ref{f2}e,
\ref{f2}f, and
\ref{f2}g
using the appropriate Feynman rules given in 
fig.\ref{f1}.  The results are given in table~\ref{t2}.

We have also provided the results of computing
the superfield diagrams of figs.\ref{f2}n and \ref{f2}o.  We 
note that we could have examined the renormalization of the coupling $\lambda$
by computing the divergent part of the six-point diagrams rather than the 
four-point diagrams in fig.\ref{f2}; 
supersymmetry ensures that the results of this approach
would yield an identical renormalization of $\lambda$.

A third model will now be considered.  This will be
done in Minkowski space and will incorporate two complex
two-component spinor fields $\psi_1$ and $\psi_2$ which form
the upper and lower components of a four component spinor field
$\psi$, as well as a pair of real scalar fields $A$ and $B$.  The
$2 \times 2$ Dirac matrices $\gamma_a$ in three dimensions,
which are discussed in the appendix, are unitarily inequivalent
to the matrices $-\gamma^a$ (which of course satisfy the same
Dirac algebra); the Dirac matrices we will use in this model are given
by 
\begin{equation}
\Gamma^a = \left( \begin{array}{cc} \gamma^a & 0 \\ 0 & -\gamma^a \end{array}
\right) \, , \;\; (a = 0,1,2 ) \, .\label{e5} 
\end{equation}
We define two additional anticommuting Dirac matrices
\begin{equation}
\Gamma_4 = \left( \begin{array}{cc} 0 & -i\openone \\ i\openone & 0 
\end{array} \right)
\;\;\; \text{and} \;\;\;\;
\Gamma_5 = \left( \begin{array}{cc} 0 & \openone \\ \openone & 0 \end{array} 
\right) \, .
\end{equation}%
The action which we will consider is initially expressed in the form
\begin{eqnarray}
S & = & \int d^3x \Big[ - \case{1}{2} \left( A p^2 A + B p^2 B \right)
- \overline \psi p \!\! / \psi + \case{g}{2} \overline \psi \psi (A^2 - B^2) 
\label{e7} \\
&  & \hspace{0.5cm} -iG\overline\psi\Gamma_5\psi AB - \case{\lambda}{6!}
(A^6 + B^6) - \case{\sigma}{4! 2!} \left(A^2B^4 + A^4 B^2 \right) \Big] 
\nonumber 
\end{eqnarray}
where $p = i \partial$ and $\overline \psi = \psi^+ \Gamma_0$.  There are 
two discrete symmetries present in this Lagrangian,
\begin{mathletters}
\begin{equation}
\label{e8a}
A \rightarrow B \, , \;\;\; B \rightarrow -A \, , 
\;\;\; \psi \rightarrow \Gamma_5 \psi \, ,
\end{equation}
and
\begin{equation}
A \rightarrow B \, , \;\;\; B \rightarrow A \, , 
\;\;\; \psi \rightarrow \Gamma_4 \psi \, . 
\end{equation}%
\end{mathletters}%
These symmetries exclude couplings of the form $\overline \psi \psi (A^2 + B^2)$,
$\overline \psi \psi AB$, $\overline \psi \Gamma_5 \psi (A^2 \pm B^2)$,
$\overline \psi \Gamma_4 \psi AB$, $\overline \psi \Gamma_4 \psi (A^2 \pm B^2)$,
$A^5 B \pm A B^5$, $A^3 B^3$, $A^2 B^4 - A^4 B^2$ and $A^6 - B^6$.
They do not exclude $\overline \psi \Gamma_4 \Gamma_5 \psi (A^2 + B^2)$, 
but this interaction is not perturbatively
generated by radiative corrections
induced by the action of (\ref{e7}).

If in (\ref{e7}), we were to set $G=g$ and $\sigma = \lambda/5$,
then the action would become
\begin{equation}
S = \int d^3 x \left[ - \phi^* p^2 \phi - \overline \psi p \!\! / \psi
+ g \overline \psi \left( \phi^2 P_+ + (\phi^*)^2 P_-\right) \psi
- \case{\lambda}{90} \left( \phi^* \phi \right) ^3 \right]
\label{e9}
\end{equation}
where $\phi = (A + i B) / {\sqrt{2}} $ and $P_\pm = (1 \pm \Gamma_5)/2$.
The symmetry of (\ref{e8a}) now becomes continuous:
\begin{equation}
\label{e10}
\phi \rightarrow e^{i \theta} \phi \, , 
\;\;\;\; \psi \rightarrow e^{-i \theta \Gamma_5/2} \psi
\end{equation}
as $P_\pm e^{-i \theta \Gamma_5/2} = P_\pm e^{\mp i \theta/2}$.

\begin{figure}
\centering
\begin{tabular}{c@{\hspace{1.5cm}}c}
\epsfig{file=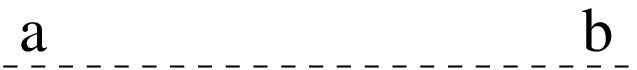, width=1.5in} & $-i\delta^{ab}/p^2$ \\
\epsfig{file=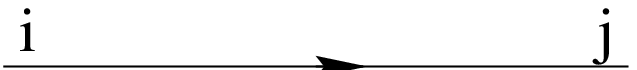, width=1.5in} &
					   $-i \delta^{ij}/p\!\!\!/ $ \\
\epsfig{file=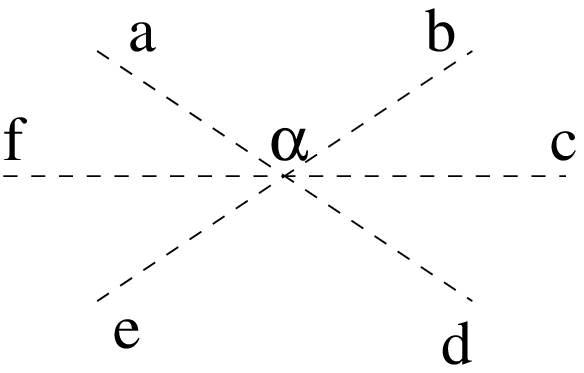,width=1.5in} & 
	\raisebox{1cm}{$-i \zeta_\alpha (\rho_\alpha \otimes \rho_\alpha
	\otimes \openone)_{[abcdef]}$} \\
\epsfig{file=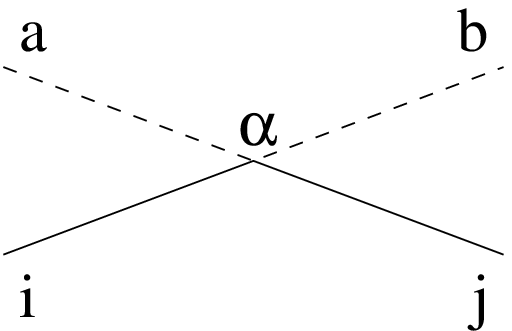,width=1.5in} & 
	\raisebox{1cm}{$i \kappa_\alpha 
	(\rho_\alpha)_{ij} (\rho_\alpha)_{ab}$}
\end{tabular}
\caption{Feynman rules for the two-component model\label{f3}}
\end{figure}

The interaction terms of (\ref{e7}) are invariant under the transformations
\begin{equation}
A = (A' + B') / {\sqrt{2}} \, , \;\;\; B = (A' - B') / {\sqrt{2}} \, , \;\;\;
\psi = (i \Gamma_5)^\frac{1}{2} \psi' = 
\frac{1 + i \Gamma_5}{\sqrt{2}} \psi'\, ,
\label{e11}
\end{equation}
provided that we simultaneously redefine the couplings so that 
\begin{mathletters}
\label{e12}
\begin{eqnarray}
g' = G & , & \;\;\; G' = g \, ,  \\
\sigma' = \frac{\lambda - \sigma}{4} &  ,&  \;\;\; 
\lambda' = (15 \sigma + \lambda) /4  \, .
\end{eqnarray}
\end{mathletters}%
By exploiting the symmetries of (\ref{e12}), we can determine how
$G$ and $\sigma$ are renormalized to a given order in perturbation
theory from results which give how $g$ and $\lambda$ are 
renormalized.  To do this directly we rewrite (\ref{e7}) in the 
form
\begin{eqnarray}
S & = & \int d^3x \left[ -\case{1}{2} \Phi^T p^2 \Phi - \overline \psi_i
p \!\! / \psi_i + \case{1}{2} \sum^2_{\alpha=1} \kappa_\alpha \left(
\Phi^T \rho_\alpha \Phi \right) \left( \overline \psi \rho_\alpha \psi \right)
\right.
\nonumber \\
& & \hspace{0.5cm} \left. - \case{1}{6!} \sum^2_{\alpha = 1} \zeta_\alpha
\left( \Phi^T \rho_\alpha \Phi \right)^2 \Phi^T \Phi \right]
\label{e13}
\end{eqnarray}
\addtocounter{equation}{-1}
where we have reverted to using the two-component spinors $\psi_i$ and have
defined
\begin{mathletters}
\begin{eqnarray}
\Phi^T & = & ( \begin{array}{cc} A, & B \end{array} ) \, ,\\
\kappa_\alpha & = & ( \begin{array}{cc} G, & g \end{array} ) \, ,\\
\zeta_\alpha & = & ( \begin{array}{cc} \case{15\sigma + \lambda}{4} , &
\lambda \end{array} ) \, , 
\end{eqnarray}
and
\begin{equation}
\label{e13d}
\rho_\alpha = ( \begin{array}{cc} \sigma_1, & \sigma_3 \end{array} )
\, .
\end{equation}
\end{mathletters}%
The transformations of equations~(\ref{e11}) and~(\ref{e12}) become
\begin{mathletters}
\begin{eqnarray}
\psi & \rightarrow & 
\case{1}{\sqrt{2}} \left( \rho_1 + \rho_2 \right) \psi \, , \\
\Phi & \rightarrow & 
\case{1}{\sqrt{2}} \left( \rho_1 + \rho_2 \right) \Phi \, ,
\end{eqnarray}%
\end{mathletters}%
and
\begin{equation}
\kappa_1 \leftrightarrow \kappa_2 
 \hspace{0.5cm} \zeta_1 \leftrightarrow \zeta_2 \, ,
\end{equation}
under which the action of  eq.(\ref{e13}) is invariant.  The 
Feynman rules associated with the action in eq.(\ref{e13}) are given 
in fig.\ref{f3}.

In table~\ref{t3} we provide the results of computing the relevant
diagrams in fig.\ref{f2} in the context of the action of eq.(\ref{e13}).
We have verified that these results can be obtained by using the action
of eq.(\ref{e7}) directly, and also that the model of eq.(\ref{e9}) yields
results that coincide with those following from eqs.(\ref{e7}) and (\ref{e13})
in the limit $g = G$, $\sigma = \lambda/5$.  

We are now in a position to use these results to determine the renormalization
group functions associated with these models to lowest order.

\section{The renormalization group functions}

The pole parts of the diagrams in fig.\ref{f2} as tabulated
in tables~\ref{t1}-\ref{t3} serve to fix the relationship
between the renormalized and bare couplings, and hence the renormalization
group functions, in the three models which we have considered 
\cite{r1,r2,r3,r4,r5}.  In other words, we can determine the functions
$a^\lambda_\nu$ and $a^g_\nu$ in the equations
\begin{mathletters}
\label{e16}
\begin{eqnarray}
\lambda_B & = & \mu^{2 \epsilon} \left[ \lambda_R + \sum^{\infty}_{\nu = 1}
\frac{a^\lambda_\nu(\lambda_R, g_R )}{\epsilon^\nu} \right] \\
g_B & = & \mu^\epsilon \left[ g_R + \sum^\infty_{\nu = 1} \frac{a^g_\nu (\lambda_R, g_R )}{\epsilon^\nu} \right] 
\end{eqnarray}
\end{mathletters}%
where, for concreteness we are using the model of eq.(\ref{e1}).  Similarly,
the wave-function renormalizations are given by
\begin{mathletters}
\begin{equation}
Z_\phi  =  1 + \sum^{\infty}_{\nu = 1} \frac{c^\phi_\nu(\lambda_R, g_R)}
{\epsilon^\nu} 
\end{equation}
and
\begin{equation}
Z_\psi = 1 + \sum^{\infty}_{\nu=1} \frac{c^\psi_\nu(\lambda_R, g_R)}{\epsilon^\nu} \, .
\end{equation}
\end{mathletters}%
Altering the mass scale $\mu$ so that 
\begin{equation}
\label{e18}
\mu' = \mu (1 + \rho) \hspace{1cm} (\rho \approx 0 )
\end{equation}
leads to expressions for $\lambda_B$ and $g_B$ that contain contributions
linear in $\epsilon$ which can be eliminated by setting
\begin{mathletters}
\label{e19}
\begin{eqnarray}
{\tilde \lambda}_R & = & \lambda_R (1- 2 \epsilon \rho) \\
{\tilde g}_R & = & g_R (1-\epsilon \rho) \, .
\end{eqnarray}
\end{mathletters}%
If we now write $\lambda_B$ and $g_B$ in terms of ${\tilde \lambda}_R$
and ${\tilde g}_R$ and identify $\lambda'_R$ and $g'_R$ (the renormalized
couplings at scale $\mu'$) with terms 
independent of poles in  $\epsilon$, we find that
\begin{mathletters}
\begin{eqnarray}
\lambda'_R & = & 
{\tilde \lambda}_R - 2 \rho a^\lambda_1 + 2 \rho {\tilde \lambda}_R
\frac{\partial a^\lambda_1}{\partial {\tilde \lambda}_R } 
+ \rho {\tilde g}_R \frac{ \partial a^\lambda_1}{\partial {\tilde g}_R} \, ,\\
g'_R & = & 
{\tilde g}_R - \rho a^g_1 + 2 \rho {\tilde \lambda}_R
\frac{\partial a^g_1}{\partial {\tilde \lambda}_R } 
+ \rho {\tilde g}_R \frac{ \partial a^g_1}{\partial {\tilde g}_R} \, ,
\end{eqnarray}
\end{mathletters}%
so that
\begin{mathletters}
\label{e21}
\begin{eqnarray}
\mu \frac{\partial \lambda}{\partial \mu} & = & -2 a^\lambda_1 +2
\lambda \frac{\partial
a^\lambda_1}{\partial \lambda} + g \frac{\partial a^\lambda_1}{\partial g} 
\, ,\\
\mu \frac{\partial g}{\partial \mu} & = & - a^g_1 + 2\lambda\frac{\partial
a^g_1}{\partial \lambda} + g \frac{\partial a^g_1}{\partial g} \, .
\end{eqnarray}
\end{mathletters}%
In determining $a^\lambda_1$ to lowest order, one must keep in mind that
a diagram involving a self-energy on an external propagator (so that the 
diagram is one particle reducible) has its divergence shared between the 
external wave-function and the coupling constant characterizing the
vertex.  This has the effect of reducing the contribution of the
pole parts of the diagrams in figs.\ref{f2}f, \ref{f2}g, and \ref{f2}l 
to the appropriate coupling constant renormalizations by a factor
of $\frac{1}{2}$ in each of the models we are examining.

Using table~\ref{t1}, we find that for the Euclidean model of eq.(\ref{e1}),
\begin{mathletters}
\label{e22}
\begin{eqnarray}
a^\lambda_1 & = & \frac{1}{\pi^2} \left(
\frac{5 \lambda^2}{96} - \frac{\lambda g^2}{2} - \frac{45}{4} g^4 \right) 
\, ,\\
a^g_1 & = & - \frac{g^3}{8\pi^3} \, ,
\end{eqnarray}
\end{mathletters}%
while for the supersymmetric model of eqs.~(\ref{e2}) and (\ref{e3})
\begin{equation}
a_1^\lambda  =  \frac{5 \lambda^3}{48 \pi^2} \, ,
\end{equation}
as can be seen from table~\ref{t2}.  Finally, for the two component
model of eqs.~(\ref{e7}) and (\ref{e13}), table~\ref{t3} gives the result
\begin{mathletters}
\label{e24}
\begin{eqnarray}
a^g_1 & = & \frac{g}{16 \pi^2} \left( 3 g^2 - 2 G^2 \right) \, ,\\
a^G_1 & = & \frac{G}{16 \pi^2} \left( 3 G^2 - 2 g^2 \right) \, ,\\
a^\lambda_1 & = & \frac{1}{16 \pi^2} \left(
 \frac{5}{6} \lambda^2 + \frac{5}{2} \sigma^2 + 16 \lambda g^2 + 
 \lambda G^2 + 15 \sigma G^2 - 360 g^4 \right) \, ,\\
a^\sigma_1 & = & \frac{1}{16 \pi^2} \left( 3 \sigma^2 + \frac{1}{3} \sigma
 \lambda + \lambda G^2 + 15 \sigma G^2 + 24 g^4 - 96 G^4 \right) \, .
\end{eqnarray}
\end{mathletters}%
The results of eq.(\ref{e24}) are consistent with the symmetries
of eq.(\ref{e12}).

Together, eq.(\ref{e21}) and the functions $a_1$ in eqs. (\ref{e22}-\ref{e24})
show that the rate of change of any coupling to lowest order 
is given by twice the corresponding $a_1$; {\it viz.} for the Euclidean model
\begin{mathletters}
\label{e25}
\begin{eqnarray}
\mu \frac{\partial \lambda}{\partial \mu} & = & \frac{1}{\pi^2}
\left( \frac{5 \lambda^2}{48} - \lambda g^2 - \frac{45}{2} g^4 \right) 
 \label{e25a}\\
\mu \frac{\partial g}{\partial \mu} & = & \frac{- g^3}{4 \pi^2} \label{e25b}
\end{eqnarray}
\end{mathletters}%
for the supersymmetric model
\begin{equation}
\label{e26}
\mu \frac{\partial \lambda}{\partial \mu}  =  \frac{5 \lambda^3}{24 \pi^2}
\end{equation}
and for the two-component model
\begin{mathletters}
\label{e27}
\begin{eqnarray}
\mu \frac{\partial g}{\partial \mu} & = & \frac{g}{8 \pi^2}
 \left( 3 g^2 - 2 G^2 \right) \label{e27a} \, ,\\
\mu \frac{\partial G}{\partial \mu} & = & \frac{G}{8 \pi^2}
 \left( 3 G^2 - 2 g^2 \right) \label{e27b} \, ,\\
\mu \frac{\partial \lambda}{\partial \mu} & = & \frac{1}{8 \pi^2}
 \left( \frac{5}{6} \lambda^2 + \frac{5}{2} \sigma^2 + 16 \lambda g^2
 + \lambda G^2 + 15 \sigma G^2 - 360 g^4 \right) \, ,\\
\mu \frac{\partial \sigma}{\partial\mu} & = & \frac{1}{8 \pi^2} \left(
 3 \sigma^2 + \frac{1}{3} \sigma \lambda + \lambda G^2 + 15 \sigma G^2
 + 24 g^4 - 96 G^4 \right) \, .
\end{eqnarray}
\end{mathletters}%
We are now in a position to discuss the properties of these
couplings.

First of all, by (\ref{e25b}), we see that
\begin{equation}
\label{e28a}
g^2 = \frac{g_0^2}{1 + \frac{g_0^2}{2\pi^2} \ln \left( \frac{\mu}{\mu_0} \right) } \equiv \frac{2\pi^2}{\ln {\mu/\Lambda} }
\;\;\; \text{or} \;\;\; g = 0 
\end{equation}
where $\Lambda$ is some renormalization invariant scale and $g=g_0$ 
when $\mu = \mu_0$.
Consequentially, the theory is
asymptotically free, a property shared by non-Abelian gauge theory in
four dimensions \cite{r6}, $\phi^3$ theory in six dimensions
\cite{r3} and $\phi^4$ theory in four dimensions when the coupling is of 
the ``wrong'' sign \cite{r7}.  In the latter two cases, the models are
unacceptable as they are energetically unstable.

Since equations~(\ref{e25}) are homogeneous in $\lambda$, $g^2$ and 
$\ln(\mu/\Lambda)^{-1}$ we one can solve explicitly for the RG flow in 
the $\lambda$-$g$ plane and thus obtain a full solution to the 
scaling behaviour of both couplings at lowest order.  In order to do this, we
multiply (\ref{e25b}) by $g$ and then divide by (\ref{e25a}) to obtain the ODE
\begin{equation}
\frac{d g^2}{d\lambda}  =  
- \frac{1}{2} \left[ \frac{g^4}{\frac{5 \lambda^2}{48} 
- \lambda g^2 - \frac{45}{2} g^4} \right] 
\end{equation}
Then, by upon setting $g^2 = \lambda z$, it is straightforward to solve 
for $z(\lambda)$.  Unless $g = 0$, the general solutions are
\begin{mathletters}
\label{lambdag}
\begin{equation}
\lambda  = \frac{g^2 \left( \left(g^2 \right) ^ {\sqrt{77/2}} - \kappa
\right) } { z_+ \left( g^2 \right) ^ {\sqrt{77/2}} - \kappa z_- }
\label{lambdag1} \text{  for some } \kappa \neq 0 \, , \text{   or} 
\end{equation}
and
\begin{equation}
g^2  =  \lambda z_\pm \, , 
\end{equation}
\end{mathletters}%
where $z_\pm = ( \pm \sqrt{77/2} - 1 ) / 90 $.
Notice that the generic solution (\ref{lambdag1})
appears to admit more than one phase; if $\kappa > 0$ then
$\lambda$ varies smoothly with $g$ and approaches zero in the ultraviolet 
limit.  
However if $\kappa < 0$ then $\lambda$ can branch outside of the perturbative 
regime when the denominator of (\ref{lambdag1}) is sufficiently small.
Whether this occurs will depend on the particular choice of renormalization
conditions. 

For the supersymmetric model, eq.(\ref{e26}) implies that
\begin{equation}
\lambda = \frac{\lambda_0^2}{1 - \case{5}{12 \pi^2} \lambda_0^2 
 \ln \left( \case{\mu}{\mu_0} \right) } \equiv \frac{12 \pi^2 / 5}{\ln (\Lambda/ \mu) } \, .
\end{equation}
This model is not asymptotically free, in accordance with the general
result of ref.\cite{r9}.

\begin{figure}
\centering
\epsfig{file=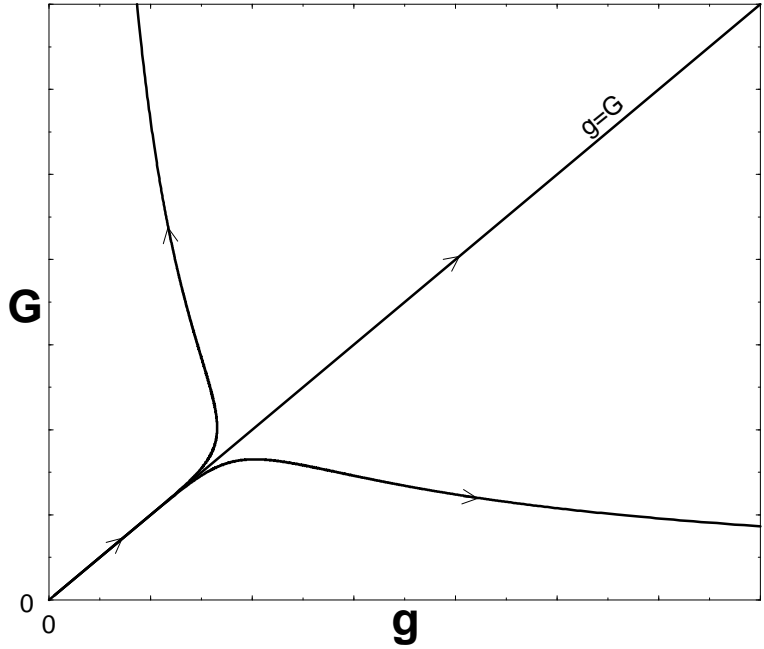,width=5in}
\caption{
Leading order flows for the Yukawa couplings in the two-component model of
equation~(\ref{e7}) 
(The arrows represent the direction of increasing energy scale.)\label{gG}}
\end{figure}

Finally, we consider the renormalization group equations for the two-component
model in (\ref{e27}).  By (\ref{e27a}) and (\ref{e27b}) we see that 
\begin{equation}
\label{e32}
\frac{d G^2}{d g^2} = \frac{3 G^4 - 2g^2 G^2} {3 g^4 - 2 g^2 G^2 }
\end{equation}
This is a homogeneous equation; upon setting $G^2 = z g^2$ it can be solved
easily to give either 
\begin{mathletters}
\label{e33}
\begin{equation}
g^2 = G^2  = \frac{4 \pi^2}{\ln(\Lambda/\mu)} \, ,
\label{e33a}
\end{equation}
(i.e. we have the Yukawa couplings of eq.(\ref{e9})) or
\begin{equation}
g^6 G^6 \propto \left| g^2 - G^2 \right| \, . \label{e33b}
\end{equation}
\end{mathletters}%
Referring to figure~\ref{gG}, the graph of eq.(\ref{e33}), we see that
the Yukawa sector has three distinct phases.  
If $G = g$, then both couplings increase with scale 
according to (\ref{e33a}) with   
their equality preserved by the renormalization group flow.  
If $g \neq G$ then equation (\ref{e33b}) holds, and  
the smaller of the two couplings 
becomes asymptotically free while the larger evolves 
outside of the perturbative region with increasing momentum scale, in such a 
way that the combination $G^4 + g^4$ always increases with $\mu$.  To see this
we set $g^2 = r \sin \theta$ and $G^2 = r \cos \theta$ in (\ref{e32}) to
obtain
\begin{equation}
\frac{dr}{d\theta} = \left[ 
\frac{3 - 5 \sin \theta \cos \theta}{\sin \theta - \cos \theta} \right]
\left[ \frac{\sin \theta + \cos \theta}{\sin\theta \cos \theta} \right] \, .
\end{equation}

Another dominant feature of figure~\ref{gG} occurs in the infrared region
where the Yukawa couplings approach zero along the asymptote $g=G$.  In fact,
we have completed several 
numerical solutions to the full four-parameter renormalization group flow 
of equations (\ref{e27}); all solutions indicate that coupling configuration
$G=g$, $\lambda = 5 \sigma$ (which leads to the chirally symmetric model
of equations (\ref{e9}) and (\ref{e10})) is asymptotically
realized in the infrared domain.  

In the special case when the Yukawa couplings $g^2$ and $G^2$ are weak,
then, in the ultraviolet limit,  
the six-point
couplings tend to branch outside of the perturbative region parallel to 
one of the two UV-stable 
lines $\lambda = \sigma$ or $\sigma = 0$.

We can slightly alter the model of eq.(\ref{e7}) so that
there are $N_g \geq 1$ Fermions coupling with strength $g$ to $A^2-B^2$,
and $N_G \geq 1$ distinct Fermions coupling with strength $G$ to $AB$.  In
this case the renormalization group equations of (\ref{e27a}) and 
(\ref{e27b})
are replaced by
\begin{mathletters}
\label{e35}
\begin{eqnarray}
\mu \frac{\partial G}{\partial \mu} & = &
\frac{G}{24 \pi^2} \left[ \left( 4 N_G + 5 \right)
G^2 - 2 N_g g^2 \right] \\
\mu \frac{\partial g}{\partial \mu} & = & \frac{g}{24 \pi^2} \left[ \left(
4 N_g + 5 \right) g^2 - 2 N_G G^2 \right]
\, .
\end{eqnarray}
\end{mathletters}%
Again we find that there are special solutions in which the couplings flow
in the following straight lines away from the origin 
as the energy scale increases:
\begin{mathletters}
\begin{eqnarray}
G = 0 \;\;\; & \text{and} & \;\;\; g^2 = \frac{12 \pi^2}
 {\left(4 N_g + 5\right) \ln (\Lambda/\mu) } \, ,\\
g  =  0 \;\;\; & \text{and} & \;\;\; G^2 = \frac{12 \pi^2}
 {\left(4 N_G + 5\right) \ln (\Lambda/\mu) } \, , 
\end{eqnarray}
or
\begin{equation}
\frac{G^2(\mu)}{g^2(\mu)}  =  \frac{6N_G + 5}{6N_g+5} \, . 
\label{eManyFa}
\end{equation} 
\end{mathletters}%
In addition there are generic solutions similar to the curves depicted in
figure \ref{gG} in which either $g$ or $G$ is asymptotically free depending
on whether the fraction $G^2/g^2$ is larger or smaller than its critical
value in (\ref{eManyFa}).

\section{Discussion}
We have considered a number of Yukawa couplings in three
dimensions as well as the associated radiatively induced six-point
scalar couplings.  The lowest order contributions to the 
renormalization group functions for the couplings 
constants in these models have been
computed, complementing the work of ref.\cite{r9}, where a global $SU(N)$
flavour symmetry is imposed on the interactions.

One curious feature of these results is that the 
Yukawa coupling is asymptotically free in the model defined
in Euclidean space by eq.(\ref{e1}), while the supersymmetric
model in Minkowski space (with metric $g_{\mu\nu} = ( + \, + \, - )$ )
whose action is given in eqs.(\ref{e2}) and (\ref{e3}) is not.
This apparent discrepancy is a consequence of the different properties of 
the Dirac
matrices $\gamma^a$ in the two models.  In Euclidean space the kinetic
term $\overline \psi p \!\! / \psi$ in eq.(\ref{e1}) is 
Hermitian provided $\gamma^a$ is identified with some unitary equivalent
representation of the Pauli spin matrices $\sigma^a$ (or $-\sigma^a$), in
which case $p \!\! / ^2 = p^2$.  In the supersymmetric
(Minkowski) model of eqs.(\ref{e2}) and (\ref{e3}), the kinetic
term for the spinor field is Hermitian provided $\gamma^a$ is represented,
for example, by eq.(\ref{eA6}), so that $p \!\! / ^2 = -p^2$.  This
point of distinction is sufficient to alter the ultraviolet 
behaviour of the models, as the integrals associated with
figs.\ref{f2}c-\ref{f2}g each contain spinor propagators.

The two component model of eq.(\ref{e7}) has the peculiar feature that unless
$g = G$, one of the couplings becomes small while the other grows as the 
renormalization group scale increases.
This differs from the analogous situation in four dimensions where none
of the couplings involving scalars and spinors is ever asymptotically free.

\section{Acknowledgements}

We wish to thank NSERC for financial assistance.  M. McInnis provided
help and encouragement.

\appendix

\section*{Conventions}

In this appendix the notation and conventions that we have been
using are outlined.  In the Euclidean model, we work in Euclidean space
with $p = -i \partial$, and the Dirac gamma matrices are identified with 
the Pauli spin matrices so that $\gamma^a \gamma^b = \delta^{ab} +
i \epsilon^{abc} \gamma^c$.
This ensures that the kinetic term
$\overline \psi = \psi^+$.  Feynman integrals are computed using 
\begin{equation}
\label{eA1}
\int \frac{d^nk}{(2 \pi)^n} \frac{(k^2)^a}{(k^2 + m^2)^b} =
\frac{1}{(4\pi)^{n/2}} \left(m^2 \right)^{n/2 + a - b}
\frac{\Gamma(n/2 + a) \Gamma(b-a-n/2)}{\Gamma(n/2) \Gamma(b)}
\end{equation}

In the supersymmetric model, the conventions of \cite{r8} are
used.  This means that an extra factor of $(i)$ appears in
eq.(\ref{eA1}), as the metric is given by $g_{\mu\nu} = ( + \, + \, - )$.
Indices are raised and lowered by the use of an antisymmetric tensor
$C_{\alpha\beta}$, so that
\begin{equation}
\label{eA2}
\psi^\alpha = C^{\alpha\beta} \psi_\beta \, , \;\;\;\; \psi_\alpha
= \psi^\beta C_{\beta\alpha}
\end{equation}
where $C_{\alpha\beta} = - C^{\alpha \beta} = \left( \begin{array}{cc}
0 & -i \\ i & 0 \end{array} \right)$.
Consequently, if ${\delta_\alpha}^\beta = \left( \begin{array}{cc}
1 & 0 \\ 0 & 1 \end{array} \right)$, then ${\delta^\alpha}_\beta =
- {\delta_\alpha}^\beta$.  
A rank-two spinor $V_{\alpha\beta}$ 
is identified with a vector 
({\it viz.} equation~(\ref{vectspin}))
provided $V_{\alpha\beta} = V_{\beta\alpha}$, 
or equivalently ${V_\alpha}^\alpha =0$.
Spinorial derivatives are defined by
$\{ \partial_\alpha, \theta^\beta \} = \{ -i p_\alpha, \theta^\beta \}
= {\delta_\alpha}^\beta$ and the covariant derivative
$D_\alpha = \partial_\alpha +i \theta^\beta\partial_{\alpha \beta}$
satisfies
\begin{eqnarray}
\left( D^2 \right) ^2 & = & \Box  \nonumber \\
\left. D^2 \delta(\theta) \right|_{\theta = 0} & = & 1 \\
\int d^3x \int d^2\theta \left[ D^2 f(\theta) \right] g(\theta) & =&
 \int d^3x \int d^2 \theta f(\theta) \left[ D^2 g(\theta) \right]
\nonumber 
\end{eqnarray}
where $D^2 = \frac{1}{2} D^\alpha D_\alpha = - \frac{1}{2} D_\alpha D^\alpha$.
We also have
\begin{equation}
\partial^{\alpha \lambda} \partial_{\beta \lambda} 
= {\delta_\beta}^\alpha \Box \, .
\end{equation}

These conventions are all consistent with replacing a vector
${V_\alpha}^\beta$ by
\begin{equation}
{V_\alpha}^\beta = {\left( \gamma^a \right)_\alpha}^\beta V_a
\label{vectspin}
\end{equation}
where $V_a$ is the usual vector in $2+1$ dimensions and 
\begin{equation}
\label{eA6}
{ \left( \gamma^a \right)_\alpha}^\beta = 
\left[ 
 \left( \begin{array}{cc} i & 0 \\ 0 & -i \end{array}\right) , \,
 \left(\begin{array}{cc} 0 & i \\ i & 0 \end{array} \right) , \,
 \left( \begin{array}{cc} 0 & i \\ -i & 0 \end{array}\right)  \right] \, .
\end{equation}
with these conventions
\begin{displaymath}
{( \gamma^a )_\alpha}^\alpha = 0 \, ,\;\;\;\;\; 
 {( \gamma^a )_\alpha}^\beta = 
 - {{( \gamma^a )_\alpha}^\beta} ^* 
\end{displaymath}
and
\begin{equation}
{(\gamma^a)_\alpha}^\lambda 
{(\gamma^b)_\lambda}^\beta 
+ 
{(\gamma^b)_\alpha}^\lambda
{(\gamma^a)_\lambda}^\beta = -2 g^{ab} {\delta_\alpha}^\beta \, .
\end{equation}

The wave equation for $\psi_\alpha$ is now consistent with
$\psi_\alpha$ being real.  The kinetic term $\psi^\alpha ( i {\partial_\alpha}
^\beta ) \psi_\beta = \psi^\alpha {(p \!\! / )_\alpha}^\beta \psi_\beta$
is Hermitian.

In the two-component model~(\ref{e13}), 
some useful identities for the matrices $\rho_\alpha$
in eq.(\ref{e13d}) are
\begin{eqnarray*}
45 \left( \rho_\alpha \otimes \rho_\alpha \otimes \openone \right)_
{[bcaaaa]} & = & 
6 \left( \rho_\alpha \right)_{bc} \left( \rho_\alpha \right) _{aa} 
\delta_{aa} + 12 \left( \rho_\alpha \right)_{ba} \left( \rho_\alpha 
\right)_{ca} \delta_{aa} \\
& & + 12 \left( \rho_\alpha \right)_{ba} \left(\rho_\alpha\right)_{aa}
\delta_{ca} + 12 (\rho_\alpha)_{ca} \left( \rho_\alpha \right)_{aa}
\delta_{ba} \\
& & + 3 \left( \rho_\alpha \right)_{aa} \left( \rho_\alpha \right)_{aa}
\delta_{bc}
\end{eqnarray*}
and
\begin{eqnarray*}
45 \left( \rho_\alpha \otimes \rho_\alpha \otimes \openone \right)_{[ bbbaaa]}
& = & 9 \left( \rho_\alpha \right)_{bb} \left( \rho_\alpha \right)_{ba}
\delta_{aa} + 9 \left( \rho_\alpha \right)_{bb} \left( \rho_\alpha \right)_{aa}
\delta_{ba} \\
& & + 18 \left( \rho_\alpha \right)_{ba} \left( \rho_\alpha \right)_{ba}
\delta_{ba}
+ 9 \left( \rho_\alpha \right)_{ba} \left( \rho_\alpha \right)_{aa}
\delta_{bb} \, .
\end{eqnarray*}
No summation is implied in these latter equations; 
square parentheses $[ \cdots ]$ indicate
symmetrization of all enclosed indices,
while repeated Roman indices $a$ and $b$ indicate
symmetrization over specific indices 
(which are thus rendered indistinguishable).  
The coefficient $45$ is shown on 
the left hand sides to elucidate the $45$ distinguishable 
permutations of the indices of $\rho_\alpha \otimes \rho_\alpha \otimes 
\openone$ 
given that both $\rho_\alpha$ and $\openone$ are already symmetric. 
 
\begin{figure}
\centering
\epsfig{file=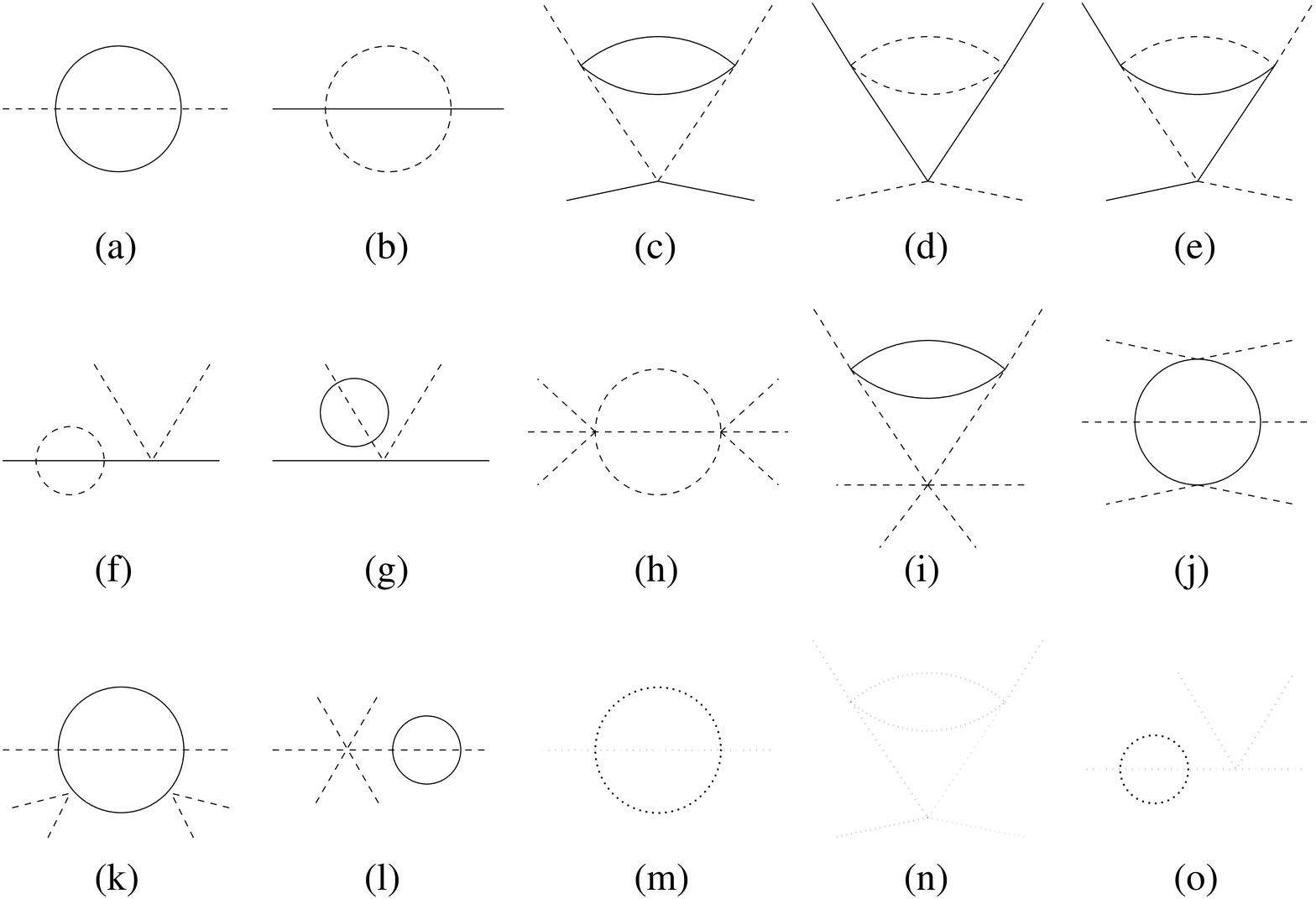, width=6.5in}
\caption{Feynman diagrams for the Euclidean, supersymmetric and two-component
models \label{f2}}
\end{figure}

\begin{figure}
\centering
\begin{tabular}{c@{\hspace{1.5cm}}c@{\hspace{1.5cm}}c}
\epsfig{file=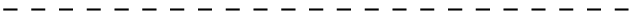, width=1.5in} & $1/p^2$ & $-i/p^2$ \\
\epsfig{file=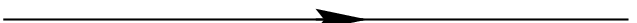, width=1.5in} & $- p \!\!\!/ / p^2 $ &
					   $-ip \!\!\!/ /p^2 $ \\
\epsfig{file=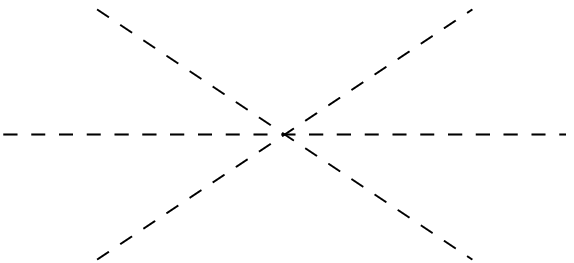,width=1.5in} & 
	\raisebox{0.25in}{$-\lambda$} & \raisebox{0.25in}{$-10i\lambda^2$}\\
\epsfig{file=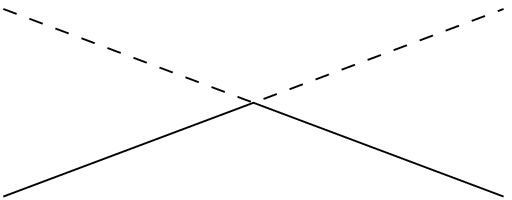,width=1.5in} & 
	\raisebox{0.25in}{$-g$} & \raisebox{0.25in}{$i\lambda$} \\
\epsfig{file=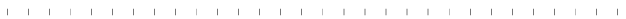, width=1.5in} & & 
	$-iD^2\delta(\theta_1 - \theta_2) /p^2$ \\
\epsfig{file=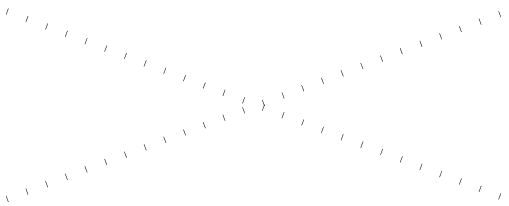, width=1.5in} & &
	\raisebox{0.25in}{$-i\lambda$} 
\end{tabular}
\caption{Feynman rules in the Euclidean and supersymmetric models \label{f1}}
\end{figure}

\begin{table}
\begin{tabular}{c|c|c|c}
Graph & Number of diagrams & Combinatoric factor & Pole Part \\ \hline
a & 1 & 1 & $g^2 p^2/96 \pi^2 \epsilon $\\
b & 1 & 1/2 & $g^2 p \!\! / /192 \pi^2 \epsilon$ \\
c & 1 & 1 & $-g^3/32\pi^2 \epsilon$ \\
d & 1 & 1/2 & $-g^3/64 \pi^2 \epsilon $\\
e & 4 & 1 & $-g^3/16 \pi^2 \epsilon$ \\
f & 2 & 1/2 & $-g^3/96 \pi^2 \epsilon$ \\
g & 2 & 1 & $-g^3/48\pi^2 \epsilon$ \\
h & 10 & 1/6 & $5 \lambda^2/96 \pi^2 \epsilon $\\
i & 15 & 1 & $-15 \lambda g^2/32 \pi^2 \epsilon$ \\
j & 90 & 1 & $-45 g^4/8\pi^2 \epsilon$ \\
k & 180 & 1 & $-45 g^4/8 \pi^2 \epsilon$ \\
l & 6 & 1 & $ - \lambda g^2 / 16 \pi^2 \epsilon$ 
\end{tabular}
\caption{Two-loop poles in the Euclidean model \label{t1}}
\end{table}

\begin{table}
\begin{tabular}{c|c|c|c}
Graph & Number of Diagrams & Combinatoric Factor & Pole Part \\ \hline
c & 1 & 1/2 & $-i \lambda^3/64 \pi^2 \epsilon $ \\
d & 1 & 1/2 & $ -i \lambda^3 / 64 \pi^2 \epsilon$ \\
e & 4 & 1 & $-i \lambda^3/16 \pi^2 \epsilon$ \\
f & 2 & 1/2 & $-i \lambda^3/96 \pi^2 \epsilon $ \\
g & 2 & 1/2 & $-i \lambda^3 / 96 \pi^2 \epsilon$ \\
n & 6 & 1/2 & $-3i \lambda^3/32 \pi^2 \epsilon$ \\
o & 4 & 1/6 & $-i \lambda^2/48 \pi^2 \epsilon$ 
\end{tabular}
\caption{Two-loop poles in the supersymmetric model \label{t2}}
\end{table}

\begin{table}
\begin{tabular}{c|c|c|c}
Graph & Number of Diagrams & Combinatoric Factor & Pole Part \\ \hline
a & 1 & 1 & $\frac{-i}{48 \pi^2 \epsilon} (\kappa_1^2 + \kappa^2_2 ) 
\openone p^2 $\\
b & 1 & 1/2 & $\frac{-i}{96 \pi^2 \epsilon} (\kappa_1^2 + \kappa^2_2 ) \openone p \!\! /$ \\
c & 1 & 1 & $\frac{-i}{16 \pi^2 \epsilon} (\kappa^3_1 - \kappa_1 \kappa_2^2 )
(\rho_1 \otimes \rho_1) + ( 1 \leftrightarrow 2)$ \\
d & 1 & 1/2 & $\frac{-i}{32 \pi^2 \epsilon} (\kappa^3_1 - \kappa_1 \kappa_2^2 )
(\rho_1 \otimes \rho_1) + ( 1 \leftrightarrow 2)$ \\
e & 4 & 1 & $\frac{-i}{16 \pi^2 \epsilon} (\kappa^3_1 - \kappa_1 \kappa_2^2 )
(\rho_1 \otimes \rho_1) + ( 1 \leftrightarrow 2)$ \\
h & 10 & 1/6 & $\frac{i}{1440 \pi^2 \epsilon} 
 (76 \zeta_1^2 - 8 \zeta_1 \zeta_2 + 16 \zeta_2^2) ( \rho_1 \otimes \rho_1 \otimes \openone ) + (1 \leftrightarrow 2) $\\
i & 15 & 1 & $\frac{15i}{16 \pi^2 \epsilon} (\zeta_1 \kappa_1^2 + \frac{4}{15} \zeta_2\kappa^2_2 - \frac{1}{15} \zeta_1 \kappa_2^2 ) ( \rho_1 \otimes \rho_1 \otimes \openone ) + (1 \leftrightarrow 2)$ \\
j & 45 & 2 & $\frac{-45i}{4 \pi^2 \epsilon} (\kappa^4_1 - \kappa_1^2 \kappa_2^2) ( \rho_1 \otimes \rho_1 \otimes \openone ) + (1 \leftrightarrow 2)$ \\
k & 45 & 4 & $\frac{-45i}{4 \pi^2 \epsilon} (\kappa^4_1 + \kappa_1^2 \kappa_2^2) ( \rho_1 \otimes \rho_1 \otimes \openone ) + (1 \leftrightarrow 2) $
\end{tabular}
\caption{Two-loop poles in the two-component model.\label{t3}}
\end{table}

\end{document}